# On the Boundary conditions for the Radial Schrodinger Equation

### Anzor A.Khelashvili and Teimuraz P. Nadareishvili

*Inst. of high Energy Physics, Iv. Javakhishvili Tbilisi State University, University Str. 9, 0109, Tbilisi, Georgia*
*and St.Andrea the First-called Georgian University of Patriarchy of Georgia, Chavchavadze Ave.53a, 0162, Tbilisi, Georgia.*

**E-mail:** teimuraz.nadareishvili@tsu.ge and anzor.khelashvili@tsu.ge

**Abstract:** Exploring the idea that equation for radial wave function must be compatible with the full Schrodinger equation, a boundary condition $u(0) = 0$ is derived.

It is well known that the radial Schrodinger equation

$$-\frac{d^2 u}{dr^2} + \frac{l(l+1)}{r^2} u - 2m(E - V(r))u = 0 \qquad (1)$$

is obtained from the full 3-dimensional Schrodinger equation

$$-\Delta \psi(\vec{r}) - 2m(E - V(r))\psi(\vec{r}) = 0 \qquad (2)$$

after the separation of variables in spherical coordinates [1, 2]

As central potentials $V(\vec{r}) = V(r)$ play an important role in physics, consequently the radial equation (1) was applied successfully in many problems.

We mention, that the transition from Cartesian to spherical coordinates is not unambiguous, because the Jacobian of this transformation $J = r^2 \sin\theta$ is singular at $\theta = n\pi (n = 0,1,2,...)$ and $r = 0$. Angular part is fixed by the requirement on continuity and uniqueness. This gives the unique spherical harmonics $Y_l^m(\theta,\varphi)$.

As regards of radial variable we remark, that in full Schrodinger equation $\vec{r} = 0$ is an ordinary point, but in the radial equation it is a point of singularity and knowledge of some boundary behavior is necessary.

Recently much attention has been devoted to the problems of Self-adjoint extension (SAE) for the inverse squared $r^{-2}$ behaved potentials in the Schrodinger equation [3]. These problems are interesting not only from academic standpoint. Numbers of physically significant quantum-mechanical problems manifest such a behavior. Hamiltonians with inverse squared like potentials appear in many systems and they have sufficiently rich physical and mathematical structures. Moreover earlier, at 60-ies of previous century singular potentials were the subject of intensive studies in connection with non-renormalizable field theoretic models. Sufficiently exhaustive reviews to singular potentials for that time are [4-6].

It turned out that there are no rigorous ways in deriving some boundary condition for the radial wave function $u(r)$ at the origin $r=0$ in case of singular potentials.

Therefore many authors content themselves by consideration only a square integrability of radial wave function don't pay attention to its behavior at the origin. Of course this is permissible mathematically and the strong theory of linear differential operators allows such an approach [7-8]. There appears so-called SAE physics, in the framework of which among physically reasonable solutions there appear also many curious results, such as bound states in case of repulsive potential [9] and so on. We think that these highly unphysical results are caused by the fact that without suitable boundary condition at the origin a functional domain for radial Schrodinger Hamiltonian is not restricted correctly [10].

Now we want to show that there is physically very transparent way to find needed restrictions.

In many classical books on quantum mechanics it is underlined that the final radial equation must be compatible with the primary full Schrodinger equation, but in our opinion this consideration is not extended to any concrete results [2,11,12]. Though discussion often beats about bush conclusions mainly are cautiously (see, e.g. book of R. Newton[13]). It seems that some significant thing is missing without deeper exploration of the idea of compatibility.

Arming with this idea let us look at derivation of the radial wave equation in more detail. Remembering that

$$\Delta(f \cdot g) = g\Delta f + 2\nabla_i f \cdot \nabla_i g + f\Delta g \tag{3}$$

where $f$ and $g$ are arbitrary two-fold differentiable functions of $r$. Let us choose

$$f = u(r) \quad \text{and} \quad g = \frac{1}{r} \tag{4}$$

This choice evidently corresponds to the commonly used representation

$$\Psi(\vec{r}) = R(r)Y_l^m(\theta,\varphi) = \frac{u(r)}{r}Y_l^m(\theta,\varphi) \tag{5}$$

Taking into account that

$$\Delta\left(\frac{1}{r}\right) = -4\pi\delta^{(3)}(\vec{r}), \tag{6}$$

we obtain

$$\Delta\left(\frac{u}{r}\right) = \frac{1}{r}\Delta u - \frac{2}{r^3}(\vec{r}\cdot\vec{\nabla})u - 4\pi\delta^{(3)}(\vec{r})u \tag{7}$$

But $\vec{r}\cdot\vec{\nabla} = r\frac{\partial}{\partial r}$ and

$$\Delta u(r) = \frac{1}{r^2}\frac{\partial}{\partial r}\left(r^2\frac{\partial}{\partial r}\right)u(r) + \frac{1}{r^2}\Delta_{\theta,\varphi}u(r) = \frac{\partial^2 u}{\partial r^2} + \frac{2}{r}\frac{\partial u}{\partial r}$$

Therefore

$$\Delta\left(\frac{u}{r}\right) = \frac{1}{r}\frac{d^2u}{dr^2} - \frac{1}{r}4\pi u(r)\delta^{(3)}(\vec{r}), \tag{8}$$

because $u(r)$ depends only on $r$ and ordinary differentiation appears instead of partial one.

After substituting into full Schrodinger equation (2), it follows equation for $u(r)$:

$$-\frac{d^2u(r)}{dr^2}+\frac{l(l+1)}{r^2}u(r)+4\pi u(r)\delta^{(3)}(\vec{r})-2m(E-V(r))u(r)=0 \qquad (9)$$

The term with 3-dimensional delta-function must be comprehended as integrated by means of $d^3\vec{r}=r^2 dr\sin\theta d\theta d\varphi$. On the other hand [14]

$$\delta^{(3)}(\vec{r})=\frac{1}{|J|}\delta(r)\delta(\varphi)\delta(\theta) \qquad (10)$$

Taking into account all above mentioned relations, one is faced to the following equation

$$-\frac{d^2u(r)}{dr^2}+\frac{l(l+1)}{r^2}u+4\pi\delta(r)u(r)-2m(E-V(r))u(r)=0 \qquad (11)$$

Comparing to the commonly used radial equation (1) we see that the extra term with $\delta(r)$ function remains. The only reasonable way to remove this undesirable term lies in the requirement

$$u(0)=0, \qquad (12)$$

if we don't want modify the Laplace operator or include compensating delta term in the potential $V(r)$ [5].

Therefore we conclude that the radial equation for $u(r)$ is compatible with the full Schrodinger equation if and only if the boundary condition $u(0)=0$ is fulfilled. *The radial equation (1) together with boundary condition (12) is equivalent to the full Schrodinger equation (2).*

Some comments are in order now: Equation for $R(r)$ has its usual form. Problem with delta function arises in the course of elimination of the first derivative term. But derivation of boundary behavior from this equation is as problematic as for $u(r)$ from Eq. (1). But now, after the (12) is established, it follows that the full wave function $R(r)$ is less singular at the origin than $r^{-1}$. It is also remarkable to note that the boundary condition (12) is valid whether potential is regular or singular. Different potentials can only determine the possible way of tending to zero of $u(r)$ at the origin.

It seems very curious that this fact was unnoticed up to now in spite of great numbers of discussion [2,5,6,11,12]. Having this boundary condition many problems can be solved by taking it into account. It is remarkable that results obtained earlier for regular potentials remain unchanged, because this boundary condition was used in such problems. On the other hand, for singular potentials this condition would have drastic influence. For example, there is the subject of discussion how to be with $r^{-l}$ solution, when $l=0$, even for regular potential. According to obtained boundary condition this term must be ignored. Many authors neglected boundary condition on the whole and were satisfied only by square integrability. But in this case after penetration to the forbidden regions and after performing a self-adjoint procedure they earn curious unphysical results.

**ACKNOWLEDGEMENTS:**

We want to thank Profs. Sasha Kvinikhidze and Parmen Margvelashvili for valuable discussions.